\documentclass[reprint,twocolumn,superscriptaddress,showpacs,nofootinbib,notitlepage]{revtex4-2}

\usepackage{graphicx}
\usepackage{latexsym,amsmath,amssymb,lmodern,float,url}
\usepackage{natbib}
\usepackage{xcolor}
\usepackage{microtype}

\usepackage[colorlinks=true,backref=false, linktocpage=true,
citecolor=blue,urlcolor=blue,linkcolor=blue,pdfpagemode=UseOutlines]{hyperref}

\def\Re {\operatorname{{Re}}}
\def\Im {\operatorname{{Im}}}
\def\Tr {\operatorname{{Tr}}}

\begin{document}
\title{Lattice Scalar Field Theory At Complex Coupling}

\author{Scott Lawrence}
\email{scott.lawrence-1@colorado.edu}
\affiliation{Department of Physics, University of Colorado, Boulder, CO 80309, USA}
\author{Hyunwoo Oh}
\email{hyunwooh@umd.edu}
\affiliation{Department of Physics, University of Maryland, College Park, MD 20742, USA}
\author{Yukari Yamauchi}
\email{yyukari@umd.edu}
\affiliation{Department of Physics, University of Maryland, College Park, MD 20742, USA}
\date{\today}

\begin{abstract}
Lattice scalar field theories encounter a sign problem when the coupling constant is complex. This is a close cousin of the real-time sign problems that afflict the lattice Schwinger-Keldysh formalism, and a more distant relative of the fermion sign problem that plagues calculations of QCD at finite density. We demonstrate the methods of complex normalizing flows and contour deformations on scalar fields in $0+1$ and $1+1$ dimensions, respectively. In both cases, intractable sign problems are readily bypassed. These methods extend to negative couplings, where the partition function can be defined only by analytic continuation. Finally, we examine the location of partition function zeros, and discuss their relation to the performance of these algorithms.
\end{abstract}

\maketitle

\section{Introduction}
Lattice field theory has proven an important approach to the study of non-perturbative quantum field theories. However, lattice Monte Carlo calculations are obstructed in various regimes by sign problems. The most famous example of these difficulties is the fermion sign problem, which occurs in theories of relativistic fermions at finite fermion density; or, relatedly, in the Hubbard model away from half-filling. Numerically more severe sign problems can be obtained in bosonic lattice models when the Euclidean action is allowed to be complex. This occurs most naturally for the Schwinger-Keldysh formalism, a method of obtaining real-time separated observables on the lattice~\cite{Alexandru:2016gsd,Alexandru:2017lqr}. A similar sign problem can be obtained by remaining at pure imaginary time, but allowing the mass and coupling of the field to be complex.

When the mass-squared and coupling are purely imaginary, the sign problem is ``infinitely bad'': to be precise, a Monte Carlo calculation will never converge. However, this is in some sense not the worst possible case. The physical theory can be analytically continued to negative real values of the coupling constant. All physical observables remain well defined (at least on the lattice), but the original expression for the partition function is no longer a convergent integral.

Historically, the study of analytically continued field theories~\cite{Witten:2010cx} motivated the idea of performing a path integral along the Lefschetz thimbles---that is, a particular integration contour in complexified field space~\cite{Cristoforetti:2012su}. This method results in an alleviated sign problem. Later, it was found that integrating along some contour that approximates the Lefschetz thimbles also yields a strong improvement in the sign problem~\cite{Alexandru:2015xva,Alexandru:2015sua,Alexandru:2016gsd,Alexandru:2017lqr}, and can even be superior to the thimbles themselves~\cite{Lawrence:2018mve}. See~\cite{Alexandru:2020wrj} for a recent review of these methods.

Integration on or near the Lefschetz thimbles is numerically difficult, due in part to the numerical expense of evolving a differential equation at every Monte Carlo step. To accelerate the process, it was proposed to train a neural network to approximate either the thimbles themselves or the aforementioned approximation~\cite{Alexandru:2017czx}. Later developments did away entirely with the reference to Lefschetz thimbles, noting that the choice of manifold could be directly optimized to minimize the severity of the sign problem~\cite{Mori:2017pne,Alexandru:2018fqp,Giordano:2022miv}.

Contour deformations are closely related to \emph{complex normalizing flows}~\cite{Lawrence:2021izu}, which are particular maps from a real parametrizing space to a complex field space. Under a complex normalizing flow, a Gaussian measure on the real space induces a desired physical Boltzmann factor on the field space.
Normalizing flows have recently applied, in the context of sign-problem-free lattice field theories~\cite{Hackett:2021idh,Kanwar:2020xzo,Albergo:2019eim,Albergo:2021bna}, as a technique for accelerating the Monte Carlo sampling of field configurations. When complex normalizing flows are considered, much of the same technology developed in that context is directly applicable to the problem of removing a sign problem.

In this paper, we demonstrate the use of optimized contour deformations and complex normalizing flows in studying a simple proving ground for bosonic sign problems: lattice scalar field theory with a complex $\phi^4$ coupling constant. We begin in Section~\ref{sec:algorithm} by defining contour deformations and complex normalizing flows, and detailing the algorithms used to train them. Section~\ref{sec:aho} applies these methods to the quantum anharmonic oscillator, at complex coupling. For this model, numerical diagonalization of the Hamiltonian is practical, and so the results of complex normalizing flows can be compared to an exact computation. Optimized contour deformations are used directly in Section~\ref{sec:2d} to study two-dimensional scalar field theory. The analytic behavior of the lattice partition function---in particular, the location of zeros and their relevance to the (non)existence of perfect contours---is discussed in Section~\ref{sec:zeros}. Finally, Section~\ref{sec:discussion} concludes, including a discussion of the various trade-offs between the two methods studied here.

All code used in the simulations discussed in this paper is available online~\cite{code}.

\section{Algorithms}\label{sec:algorithm}
This paper uses two closely related algorithms for alleviating or removing sign problems associated to lattice computations with complex actions. Both stem from the observation that the domain of integration of the path integral can be deformed into complex space without changing the physics of the model. The first algorithm is the ``sign-optimized manifold'' approach, introduced in \cite{Mori:2017pne,Alexandru:2018fqp} and  since applied to a variety of models. The second algorithm, based on training complex normalizing flows, was suggested in~\cite{Lawrence:2021izu} (the same paper that introduced complex normalizing flows); the current work represents the first numerical demonstration of machine learning of complex normalizing flows.

Both algorithms build on the technique of ``reweighting'': a straightforward but inefficient way to deal with any sign problem. Expectation values with respect to a complex action $S(\phi)$ may be expressed in terms of a ratio of \emph{quenched} expectation values $\langle \cdot \rangle_Q$:
\begin{align}\label{eq:reweighting}
\langle \mathcal O\rangle
=
&\frac{\langle \mathcal O(\phi) e^{-i \Im S(\phi)}\rangle_Q}{\langle e^{-i \Im S(\phi)}\rangle_Q}\\
&\text{where }
\langle f(\phi)\rangle_Q \equiv \frac{\int \mathcal D\phi\, f(\phi) e^{-\Re S(\phi)}}{\int \mathcal D\phi\,e^{-\Re S(\phi)}}
\nonumber\text.
\end{align}
Computing this ratio of quenched expectation values to within a fixed precision typically incurs an exponential cost in the volume of the system---either the physical volume, as in the case of the finite-fermion-density sign problem, or the lattice volume as in the case of bosonic real-time sign problems. For physically relevant system sizes, this basic strategy is thus insufficient.

\subsection{Contour deformations}\label{sec:contours}

Consider the path integral for a lattice quantum field theory of a single real scalar field, in Euclidean spacetime:
\begin{equation}
Z = \int d^V\!\phi\; e^{-S(\phi)}
\text.
\end{equation}
Here the integral is taken over $V$ real degrees of freedom. However, under mild conditions, the domain of integration can be changed---deformed into the complexified space $\mathbb C^V$---without affecting the value of $Z$.
Let $\mathcal M$ be a $V$-real-dimensional oriented submanifold of $\mathbb C^V$ such that $-\mathcal M$ (that is, $\mathcal M$ with orientation reversed) and $\mathbb R^V$ together form the boundary of a region $\Omega$: $\partial \Omega = \mathbb R^V - \mathcal M$. Furthermore, let $f(\phi)$ be a holomorphic function whose magnitude has a finite upper bound on $\Omega$. Then it follows from the multidimensional form of Cauchy's integral theorem~\cite{Lawrence:2020irw} that two integrals are equal:
\begin{equation}
\int_{\mathbb R^V} d^V\!\phi\;f(\phi) = \int_{\mathcal M} d^V\!\phi\;f(\phi)
\end{equation}
For any particular function $f(\phi)$, the requirement that $|f(\phi)|$ possesses a finite upper bound is translated into a constraint on $\Omega$: it may not contain the regions at infinity where the real part of the action $S(\phi)\equiv -\log f(\phi)$ becomes arbitrarily small.

Not only does the numerical value of the partition function not change under appropriate contour deformations, but expectation values of reasonable observables do not either. If an observable $\mathcal O$ can be written as a function of the fields $\phi$ such that $\mathcal O(\phi) e^{-S(\phi)}$ is holomorphic, then the numerator of the expectation value
\begin{equation}
\langle \mathcal O \rangle = \frac 1 Z \int_{\mathcal M} d^V\!\phi\; e^{-S(\phi)} \mathcal O(\phi)
\end{equation}
is unchanged by deformations of the integration contour $\mathcal M$ just as the denominator is. It follows that observables can be extracted from a Monte Carlo performed on any appropriately deformed contour.

The severity of a sign problem is usually measured by the magnitude of the \emph{average phase}, defined as the ratio of the partition function to the quenched partition function:
\begin{equation}\label{eq:average-phase}
\langle \sigma \rangle \equiv
\frac Z {Z_Q} \equiv 
\frac{\int d^V\!\phi\; e^{-S(\phi)}}{\int \big|d^V\!\phi\; e^{-S(\phi)}\big|}
\text.
\end{equation}
The use of the average phase a measure of the severity of the sign problem is motivated by its appearance in the denominator of Eq.~(\ref{eq:reweighting}). The average phase is always a complex number of magnitude $|\langle \phi \rangle| \le 1$. When the magnitude is near $1$, the sign problem is mild; a magnitude near zero represents a severe sign problem. Crucially, the average phase does not take the form of an integral of a holomorphic function. In particular, the denominator is a nontrivial function of the contour of integration chosen.

The above discussion suggests that we may be able to efficiently simulate systems with sign problems by choosing an appropriate contour, and performing a Monte Carlo along that contour instead of the real plane $\mathbb R^V$. To this end we must describe first how a Monte Carlo may be performed on an arbitrary manifold, and second how a ``good'' manifold might be identified.

To integrate along a contour $\mathcal M$, we first parametrize the contour by the real plane\footnote{This parametrization step excludes all contours which are not homeomorphic to $\mathbb R^V$. An example of such a contour may be obtained by taking $\Omega$ be topologically the product of $[0,1]$ and a $V$-dimensional ball, identifying the boundaries of the interval with disjoint regions on $\mathbb R^V$. Whether such a contour is ever desirable in fighting a sign problem remains an open question.}. That is, we choose a map $\tilde\phi(\phi)$ such that $\mathcal M = \tilde\phi(\mathbb R^V)$. Such a map may be taken as the definition of the contour $\mathcal M$ in the first place. We may now introduce the \emph{effective action}, which obeys
\begin{equation}\label{eq:effective-action}
e^{-S_{\mathrm{eff}}(\phi)}
=
e^{-S[\tilde\phi(\phi)]}
\det \frac{\partial \tilde\phi}{\partial \phi}
\text.
\end{equation}
The deformed path integral over $\tilde\phi \in \mathcal M$ is now equal to a path integral over $\phi \in \mathbb R^V$ which uses the effective action in place of the original, physical action:
\begin{equation}
\int_{\mathcal M} d^V\!\tilde\phi\;e^{-S(\tilde\phi)}
= \int_{\mathbb R^V} d^V\!\phi\;e^{-S_{\mathrm{eff}}(\phi)}
\text.
\end{equation}
Similar identities hold for all observables. This final expression leads directly to an algorithm for performing a Monte Carlo on $\mathcal M$: we simply perform a Monte Carlo (using Metropolis or any other method appropriate for nonlocal actions) on the real plane, and apply $\tilde\phi$ to obtain physical field configurations for computing observables.

How can an appropriate contour $\mathcal M$---or equivalently, a map $\tilde \phi$---be selected? No general method for this task is known. However, if we restrict ourselves to some finite-dimensional family of contours $\mathcal M_\lambda$, the task is more tractable: we can now perform gradient descent within this restricted space in an attempt to optimize away the sign problem. This is the fundamental idea of ``sign-optimized manifolds''~\cite{Mori:2017pne,Alexandru:2018fqp}.

Neural networks provide a way to construct a reasonable family of contours $\mathcal M_\lambda$. Many architectures are possible; in this paper, a neural network consists of a sequence of linear transformations $\phi \mapsto A \phi + b$ interspersed with nonlinear maps $\sigma$ applied to each element independently: $\phi_i \mapsto \sigma(\phi_i)$. Given two functions defined by neural networks $f(\phi)$ and $g(\phi)$, each mapping $V$ real variables to $V$ real variables, a contour can be defined by the map
\begin{equation}
\tilde\phi(\phi) = \phi + f(\phi) + i g(\phi)
\text.
\end{equation}

Finally, in order to perform gradient descent (using, in this paper, \texttt{ADAM}~\cite{kingma2014adam}), we must select a cost function. A natural choice is the real part of the logarithm of the average phase: $C(\lambda) = -\Re \log |\langle \sigma\rangle_{\mathcal M_\lambda}|$. The minimum possible value is zero, corresponding to the complete removal of a sign problem.

This cost function cannot be evaluated efficiently! When $\langle \sigma \rangle$ is small, of order $\langle \sigma\rangle ^{-2}$ samples are needed to distinguish it from zero. In physical systems, this yields exponential scaling with the volume. However, gradient descent does not use the value of the cost function; only its derivatives enter the algorithm. The derivatives of $C(\lambda)$ are readily seen to correspond to \emph{quenched} expectation values:
\begin{equation}\label{eq:gradient}
\frac{\partial}{\partial \lambda} C(\lambda)
=
\frac{\partial}{\partial\lambda} \Re\log Z_Q
=
- \frac{\int e^{-\Re S} \frac{\partial}{\partial\lambda} \Re S}{\int e^{-\Re S}}
\text.
\end{equation}
Thus these derivatives may be efficiently computed in a Monte Carlo even in the presence of a severe sign problem~\cite{Alexandru:2018fqp,Alexandru:2018ddf}.

One final complication must be considered: as discussed at the top of this section, in order for the trained contour $\mathcal M_\lambda$ to yield the same expectation values as the starting contour $\mathbb R^V$, it must form (together with that real plane) the boundary of a region $\Omega$ on which the magnitude of the integrand has a finite upper bound. In order to trust the contour optimization procedure, it is necessary that we demonstrate that it can only yield such contours. Without a carefully engineered family of neural networks, this is only rigorously true in the limit of slow training and precisely measured gradients $\partial C$. In that limit, the gradient descent procedure may be seen as tracing out the $(V+1)$-dimensional region $\Omega$, and the boundedness of quenched partition functions obtained by gradient descent is sufficient to ensure that $\Omega$ does not contain regions with unbounded Boltzmann factors.

To summarize: in the contour deformation method, a family of integration contours $\mathcal M_\lambda \subset \mathbb C^V$ is first specified, and a parametrizing map $\tilde\phi_\lambda : \mathbb R^V \rightarrow \mathcal M_\lambda$ defined for each. Next we optimize the choice of $\lambda$ by performing stochastic gradient descent on the cost function $C(\lambda)$---each step of the descent requires a short, sign problem-free Monte Carlo to measure the gradient given by Eq.~(\ref{eq:gradient}). Finally, once a suitable contour is found, a final Monte Carlo calculation is performed to measure observables via reweighting.

\subsection{Complex normalizing flows}\label{sec:flows}
A normalizing flow is a map $\tilde\phi(\phi)$ with the property
\begin{equation}\label{eq:nf-def}
\left(\det \frac{\partial\tilde\phi}{\partial\phi}\right)e^{-S[\tilde\phi(\phi)]} \approx \mathcal N e^{-\phi^2 / 2}
\end{equation}
for some fixed normalization constant $\mathcal N$. In other words, given normally distributed $\phi$, the map $\phi\rightarrow\tilde\phi(\phi)$ induces the probability distribution $e^{-S(\tilde\phi)}$.

In theories without a sign problem, the algorithmic merit of a normalizing flow stems from the fact that it can be used to compute expectation values without the need for a Markov Chain. Independent samples $\phi$ can be efficiently obtained from the Gaussian distribution, and used to compute expectation values by applying $\tilde\phi$ to each sample. In the event that the map $\tilde\phi(\cdot)$ is merely an approximate normalizing flow (that is, Eq.~(\ref{eq:nf-def}) does not hold exactly), we define the induced action via
\begin{equation}\label{eq:nf-induced}
\left(\det \frac{\partial\tilde\phi}{\partial\phi}\right)e^{-S_{\mathrm{induced}}[\tilde\phi(\phi)]} = e^{-\phi^2 / 2}
\end{equation}
and expectation values with respect to the desired action must now be obtained by reweighting:
\begin{equation}\label{eq:nf-reweighting}
\langle\mathcal O\rangle
=
\frac{\langle \mathcal O \; e^{S_\mathrm{induced} - S}\rangle_n}{\langle e^{S_\mathrm{induced} - S}\rangle_n}
    \text.
\end{equation}
Above, $\langle\cdot\rangle_n$ denotes an expectation value taken with respect to the normal distribution over $\phi$.

Normalizing flows have recently emerged as a practical tool for accelerating Monte Carlo simulations of lattice field theories~\cite{Hackett:2021idh,Kanwar:2020xzo}. In~\cite{Lawrence:2021izu}, it was observed that when the action $S$ is permitted to be complex, the defining equation of normalizing flows, Eq.~(\ref{eq:nf-def}), also defines a parametrization of a deformed contour on which there is no sign problem. As a result, attempting to train a normalizing flow in the presence of a sign problem leads to an algorithm that has much in common with previously studied contour deformation-based methods.

In the real case (i.e.~in the absence of a sign problem), normalizing flows are known to always exist. The conditions under which complex normalizing flows exist remain unclear. Certainly it is the case that a complex normalizing flow exists if and only if there is a perfect integration contour on which the average phase has magnitude exactly $1$.

As in the case of contour deformations, it is convenient to parametrize (a subset of) the space of normalizing flows via neural networks. It remains to define a suitable cost function. In this paper, we will adopt a cost function with two terms, one addressing the log of the magnitude of the Boltzmann factor, and the second the phase.
\begin{widetext}
\begin{equation}
C_{\mathrm{nf}}(\lambda) =
\Bigg\langle\bigg|\underbrace{\frac{\phi^2}{2} + \Re\left[\log \det \frac{\partial\tilde\phi_\lambda}{\partial\phi} - \log\mathcal N - S(\tilde\phi_\lambda(\phi)) \right]}_{\Re (S_{\mathrm{induced}} - S)}\bigg|^2\Bigg\rangle_n
+
\Bigg\langle\bigg|1-\underbrace{\Big(\mathop{\mathrm{csgn}} \det \frac{\partial \tilde\phi_\lambda}{\partial \phi} \Big)e^{-i\Im \left(S(\tilde\phi_\lambda(\phi)) + \log \mathcal N\right)}}_{e^{i \Im \left(S_{\mathrm{induced}} - S\right)}}\bigg|^2\Bigg\rangle_n
\text.
\end{equation}
\end{widetext}
Above, $\mathop{\mathrm{csgn}} z = \frac z {|z|}$ denotes the complex signum function.
Although in principle a different relative weighting could be advantageous, we have not explored this. In training, real and imaginary parts of the  normalization $\mathcal N$ are treated as two more parameters to be trained.

Training of a complex normalizing flow is somewhat less straightforward than training either a real normalizing flow or an integration contour. In particular, it is difficult to guarantee that the trained normalizing flow has the correct asymptotic behavior for Cauchy's integral theorem to apply. The strategy of \emph{adiabatic training}, as discussed in~\cite{Hackett:2021idh}, is useful here. A one-parameter family of actions $S_\alpha$ is specified, for which $S_0$ is well-understood and a normalizing flow can be exactly prepared, and $S_1$ is the desired physical action. In small steps of $\Delta\alpha$, the action is advanced through this family, with a normalizing flow being re-trained at each step. Many options for adiabatic paths through the space of actions are available; for simplicity, in this paper we will use
\begin{equation}
S_\alpha(\tilde\phi)
=
\alpha S(\tilde\phi) + \frac{(1-\alpha)}{2}\tilde\phi^2
\text,
\end{equation}
except where otherwise noted\footnote{The attentive reader will note that this adiabatic path is not appropriate for cases where the original path integral fails to converge when taken along the real plane. Thus, a different path is required for the analytic continuation performed in Section~\ref{sec:aho}.}, and the adiabatic training is performed with $\Delta\alpha=0.05$.

Since a normalizing flow found numerically is certain to be only approximate, it is necessary to reweight according to Eq.~(\ref{eq:nf-reweighting}). When the required reweighting is large enough to be inconvenient, some performance gains can be obtained by performing importance sampling in the parametrizing space. Because the parametrization is approximately a normalizing flow, the Markov chain is likely to mix rapidly. As long as the sign problem is not severe, this represents an alternative method for extracting expectation values. In this work we will take care to train the normalizing flows precisely enough that this approach is not necessary.

Ordinarily, extracting the partition function itself is not natural with Monte Carlo methods\footnote{Although the partition function can be extracted, indeed in polynomial time, by constructing a sequence of Boltzmann factors connecting the physical Boltzmann factor to a well-understood trivial (e.g.~Gaussian) Boltzmann factor, and measuring the overlaps between adjacent theories.}. However, equipped with an exact normalizing flow, the partition is determined by the trained value of $\mathcal N$. With an approximate normalizing flow, the partition function may be determined by averaging over samples much as with any other observable:
\begin{equation}
\frac{Z}{Z_n} = \langle e^{S_{\mathrm{induced}}-S}\rangle
\text.
\end{equation}

\section{Anharmonic Oscillator}\label{sec:aho}
\begin{figure}
\centering\includegraphics[width=0.94\linewidth]{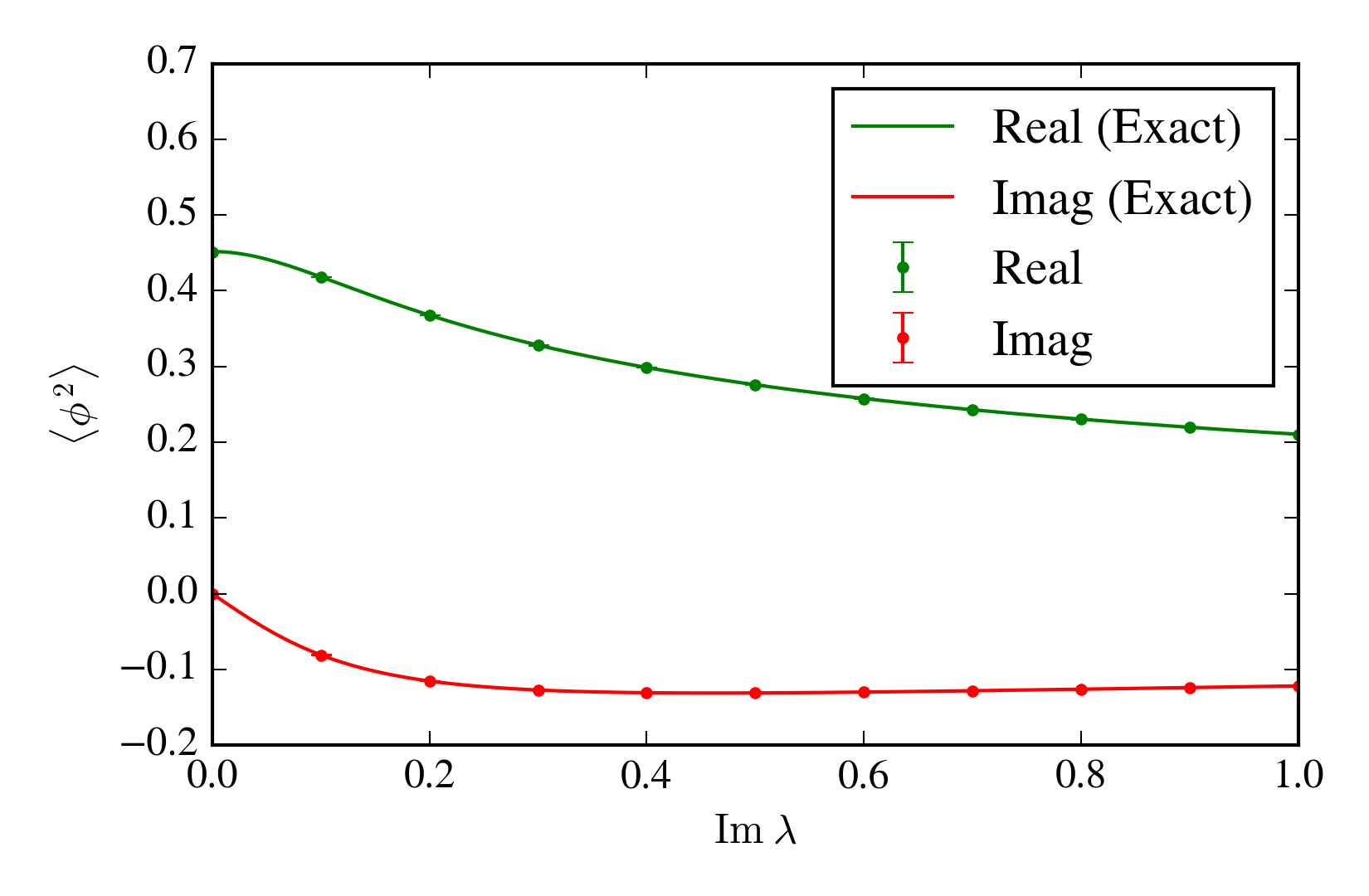}
\caption{Expectation value $\langle \phi^2\rangle$ as a function of $\Im \lambda$, with $m^2 = 0.5$ and $\Re \lambda = 0.1$, on a one-dimensional, $10$-site lattice. The exact result shown is obtained by direct diagonalization of the (non-Hermitian) Hamiltonian.\label{fig:aho}}
\end{figure}

In this section we apply the methods discussed above to study the anharmonic oscillator at complex coupling $\lambda$. The system is defined by the Hamiltonian
\begin{equation}\label{eq:H_aho}
H_{\mathrm{aho}} = \frac 1 2 p^2 + \frac{m^2}{2}x^2 + \lambda x^4
\text.
\end{equation}

The usual procedure of expanding the partition function $Z = \Tr e^{-\beta H}$ in a Trotter-Suzuki approximation yields an action defined on a one-dimensional lattice:
\begin{equation}\label{eq:S_aho}
S_{\mathrm{aho}} =
\sum_{x} \left[
\frac {(\phi_{x} - \phi_{x+1})^2} 2
+\frac{m^2}{2} \phi_{x}^2
+ \lambda \phi_{x}^4
\right]
\text,
\end{equation}
where the sum is taken over all sites, and periodic boundary conditions are assumed.

The continuum limit (in time) is obtained either by adding an explicit lattice spacing $\delta$ to the action, or equivalently by appropriately rescaling the coupling constants and fields. With $m = \delta \hat m$ and $\lambda = \delta^2 \hat \lambda$, observables of the form $\delta^\alpha \langle x^{2\alpha}\rangle$ approach a well-defined limit, given by their expectation value under the ground state of the Hamiltonian in Eq.~(\ref{eq:H_aho}).

For the purposes of numerical tests in this paper, we will not attempt to approach the continuum limit. Instead, expectation values obtained using the action of Eq.~(\ref{eq:S_aho}) are directly compared to those obtained from direct diagonalization of the Trotterized imaginary-time evolution operator.

As described in Section~\ref{sec:flows}, we construct normalizing flows $\tilde\phi(\phi)$ from neural networks. The precise functional form used is
\begin{equation}
\tilde\phi(\phi)
= (M_R + i M_I)\phi + (L_R + i L_I)f(\phi)
\text.
\end{equation}
The function $f(\phi)$ is a simple multilayer perceptron, with $\sigma(x) = \frac{1}{1 + e^{-x}}$---the sigmoid function---as the nonlinear activation function. The network has $V$ inputs, $2V$ nodes on every internal layer, and $2V$ outputs. Finally, the four matrices $M_R$, $M_I$, $L_R$, and $L_I$ are trainable parameters, as are the weights and biases of $f(\cdot)$. These matrices, as well as all parameters in $f(\cdot)$, are real.

At first glance, the separation of the $M_*$ terms from the definition of $f(\cdot)$ is unnecessary and redundant. However, note that because of the choice of activation function, the asymptotic behavior of $f(\cdot)$ is constrained. In particular, for any fixed set of parameters, $f(\phi)$ is bounded from both above and below. As a result, a normalizing flow constructed from $f(\cdot)$ alone can never map the real plane to a contour satisfying the conditions of Cauchy's integral theorem. The simple linear terms restore this capability in a straightforward way.

To check our approach and code, we show in Figure~\ref{fig:aho} the expectation value $\langle \phi^2\rangle$ for various complex $\lambda$. The exact line is computed via direct matrix manipulations of the Trotterized imaginary-time evolution operator. These exact results show excellent agreement with the Monte Carlo computations.

\begin{figure}
\centering\includegraphics[width=0.94\linewidth]{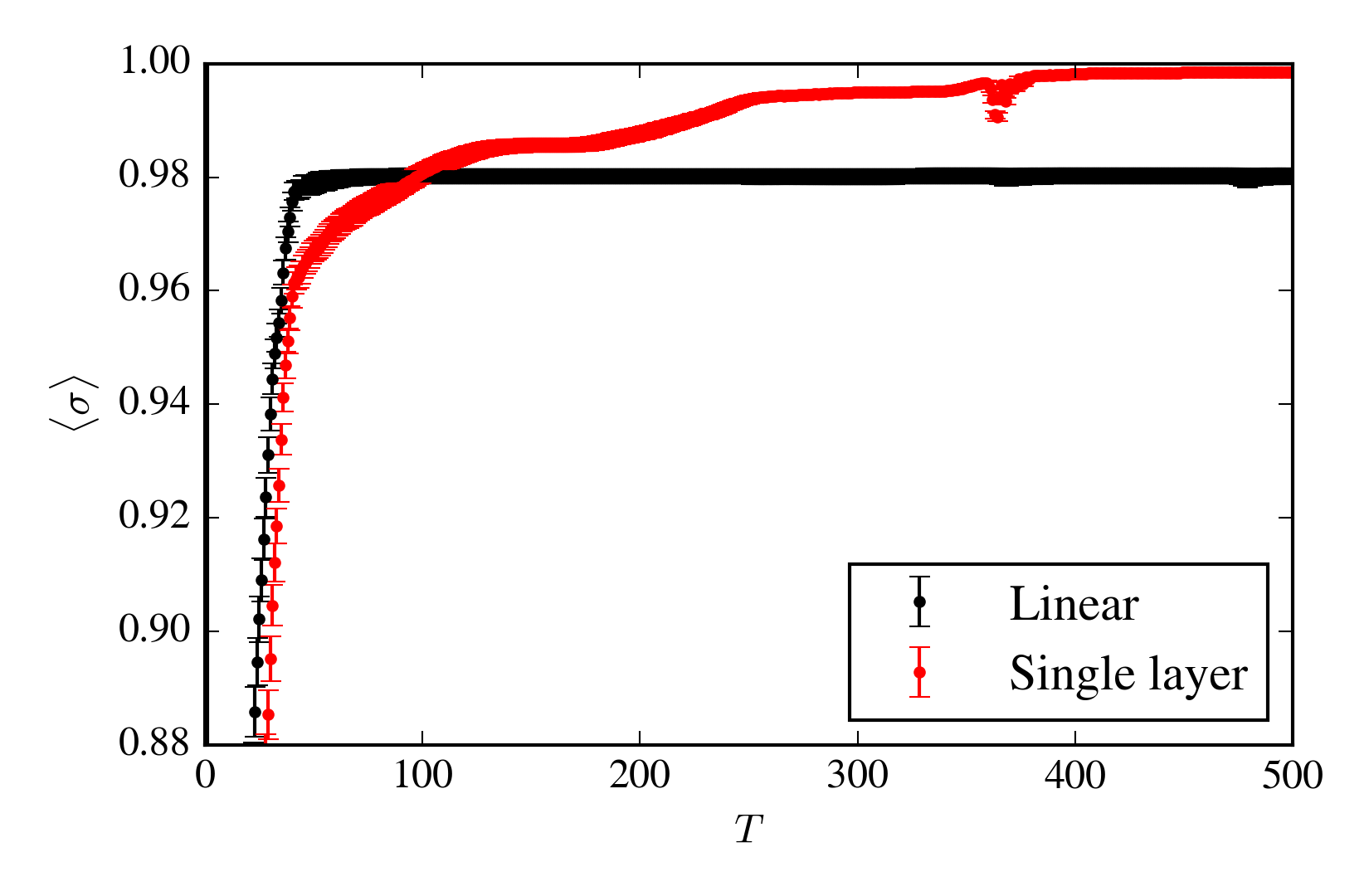}
\caption{\label{fig:perfect}The average phase achieved, as a function of training step, for two different topologies of neural network. The model used is a $10$-site lattice with $m^2=0.5$ and $\lambda = 0.1+i$. In both cases, networks are trained with \texttt{ADAM}'s standard parameters and a learning rate of $10^{-4}$. Each training step is based on $2^{15}$ samples from the Gaussian distribution.}
\end{figure}

After a normalizing flow has been trained adiabatically, it is generally advantageous to perform a more careful training with the final parameters. This ``refining'' step makes the subsequent sampling more efficient. The performance of the training process, measured by average phase, for two different topologies of network is shown in Figure~\ref{fig:perfect}.

In~\cite{Lawrence:2021izu} it was argued that perfect integration contours---that is, contours along which the sign problem vanishes entirely---are likely to exist for most values of the couplings $m^2$ and $\lambda$. As discussed in that paper, this is equivalent to the existence of exact normalizing flows. Figure~\ref{fig:perfect} may be interpreted as giving weak numerical evidence that a perfect contour indeed exists for a $10$-site lattice with $m^2=0.5$ and $\lambda = 0.1+i$. As we will discuss below, this is not true for all values of $\lambda$.

The theory described by the action Eq.~(\ref{eq:S_aho}) is initially defined only for $\Re \lambda \ge 0$. For negative $\Re \lambda$, the path integral $\int \mathcal D\phi\,e^{-S}$ diverges due to the unstable $\phi^4$ term---equivalently, the Hamiltonian Eq.~(\ref{eq:H_aho}) is unbounded below. However, the theory can nevertheless be analytically continued to values of $\lambda$ outside the region of convergence of the initial path integral. The resulting function is analytic away from the branch point at $\lambda = 0$.

A useful way to think about this analytic continuation is as a rotation of the field variables $\phi \rightarrow e^{i \theta} \phi$. Under such a rotation, the coupling transforms via $\lambda \mapsto e^{-4 i \theta} \lambda$; thus for any coupling we can select $\theta$ to make the path integral convergent, at the cost of introducing phases on the $m^2$ and $\partial^2$ terms.

\begin{figure}
\centering\includegraphics[width=0.94\linewidth]{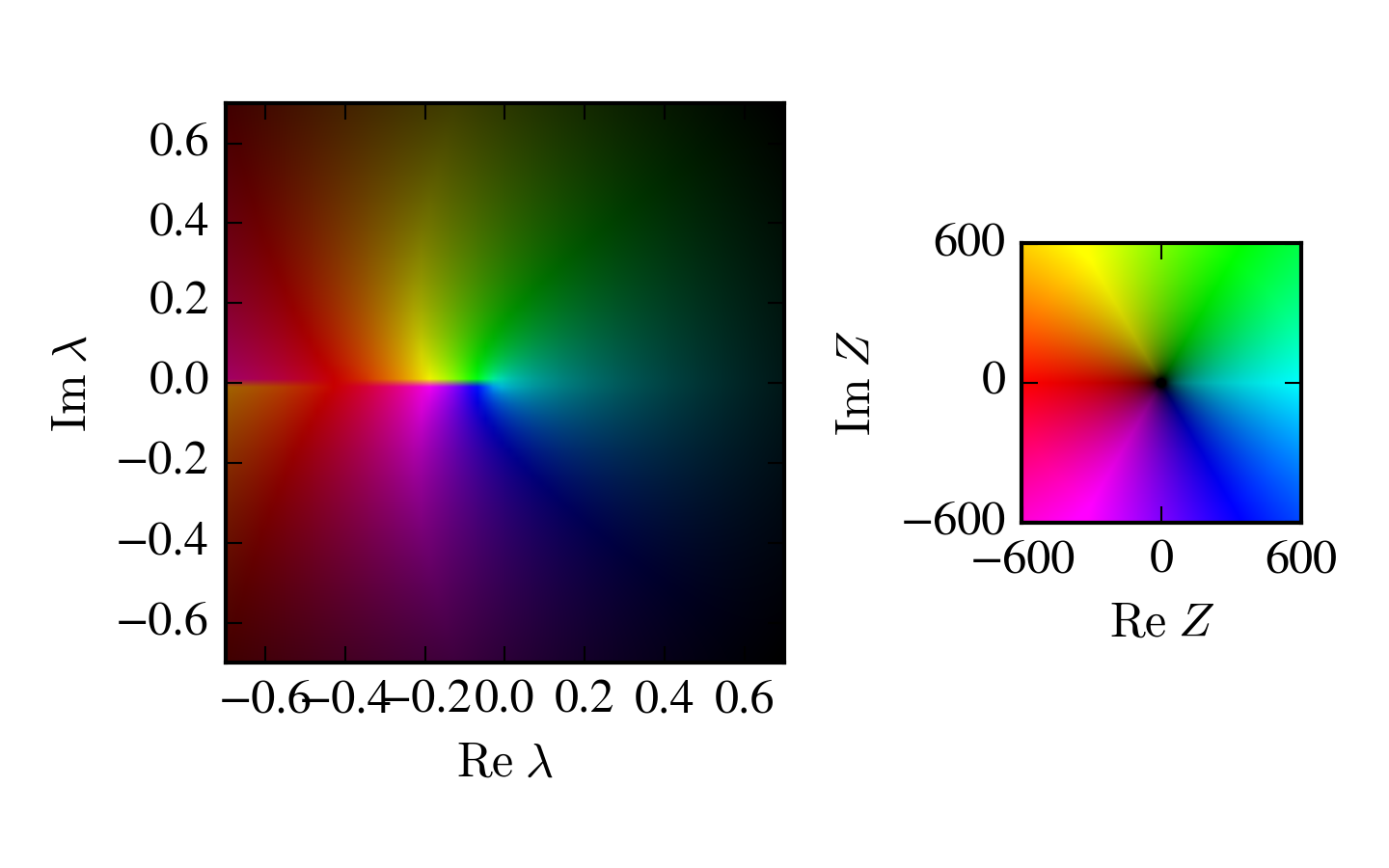}
\caption{\label{fig:partition}The partition function for lattice scalar field theory in $0+1$ dimensions, with $10$ lattice sites and $m^2=0.5$, at complex values of the coupling $\lambda$. The partition function at negative $\lambda$ is defined via analytic continuation with a branch cut placed along the negative real axis. The color scheme is shown at right, with a plot of $f(\lambda) = \lambda$ on the same scale.}
\end{figure}

As discussed in Section~\ref{sec:flows}, having a normalizing flow enables the direct computation of the partition function. This computation is shown in Figure~\ref{fig:partition}. In this demonstration, a neural network with one internal layer was used for normalizing flows. The procedure for adiabatically training normalizing flows is naturally blind to the question of whether the path integral is convergent on the original (real) domain of integration. Instead, it is equivalent to performing analytic continuation along the chosen adiabatic path. To produce the normalizing flows used for Figure~\ref{fig:partition}, flows were first prepared at various radii ($0.1,0.2,\ldots,1.0$) along the real-$\lambda$ axis, and then adiabatically retrained while rotating $\lambda$ in the complex plane. The process was limited to angles in $[-\pi,\pi]$, resulting in the branch cut apparent on the negative $\lambda$ axis.

\section{Two Dimensions}\label{sec:2d}
In this section we switch to studying two-dimensional scalar field theory on an isotropic lattice with periodic boundary conditions. The action defining this model is
\begin{widetext}
\begin{equation}
S_{\mathrm{sft}} =
\sum_{x,y} \left[
\frac {(\phi_{x,y} - \phi_{x+1,y})^2} 2
+\frac {(\phi_{x,y} - \phi_{x,y+1})^2} 2
+\frac{m^2}{2} \phi_{x,y}^2
+ \lambda \phi_{x,y}^4
\right]
\text,
\end{equation}
\end{widetext}
where $m^2$ and $\lambda$ are again permitted to be arbitrary complex numbers.

\begin{figure}
\centering\includegraphics[width=0.94\linewidth]{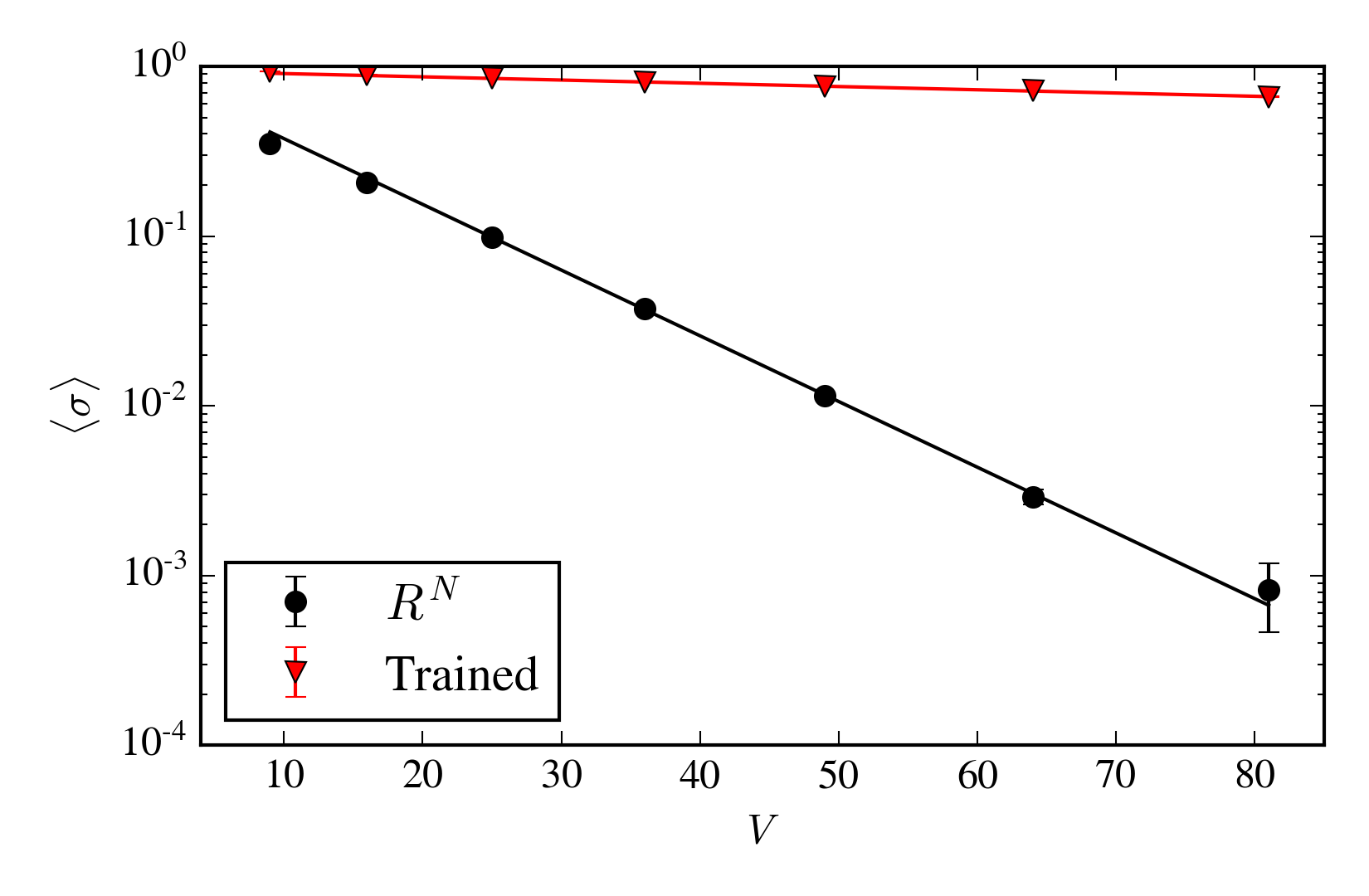}
\caption{The average sign as a function of volume, for two-dimensional scalar field theory with $m^2=0.5$ and $\lambda=1.0i$. The trained contour is constructed from a zero-layer network; that is, the integration contour is parametrized by a linear function. The numbers of configurations are $5\times10^6$ on the real plane and $5\times10^5$ on the trained contours, where fewer configurations are needed due to the improved average sign. (Statistical error bars by bootstrapping with blocking are shown, but fit within plotted points.)\label{fig:2d-sign}}
\end{figure}

To study this theory, we train integration contours following the algorithm described in Section~\ref{sec:contours}. For the moderate lattice sizes simulated, linear contours described by
\begin{equation}
\tilde\phi(\phi) = \phi + M_R \phi + i M_I \phi
\end{equation}
are found to sufficiently control the sign problem. Here $M_R$ and $M_I$ are $V\times V$ real matrices treated as the trainable parameters, with $V$ the number of sites on the lattice.
Figure~\ref{fig:2d-sign} shows the performance of integration contours trained in this way as a function of the lattice size, for lattice sizes $3\times 3$ through $9\times 9$. We used a zero-layer network for training contours. The network is fully connected and there are no additional methods used.

\begin{figure*}
\begin{center}
\includegraphics[width=0.45\linewidth]{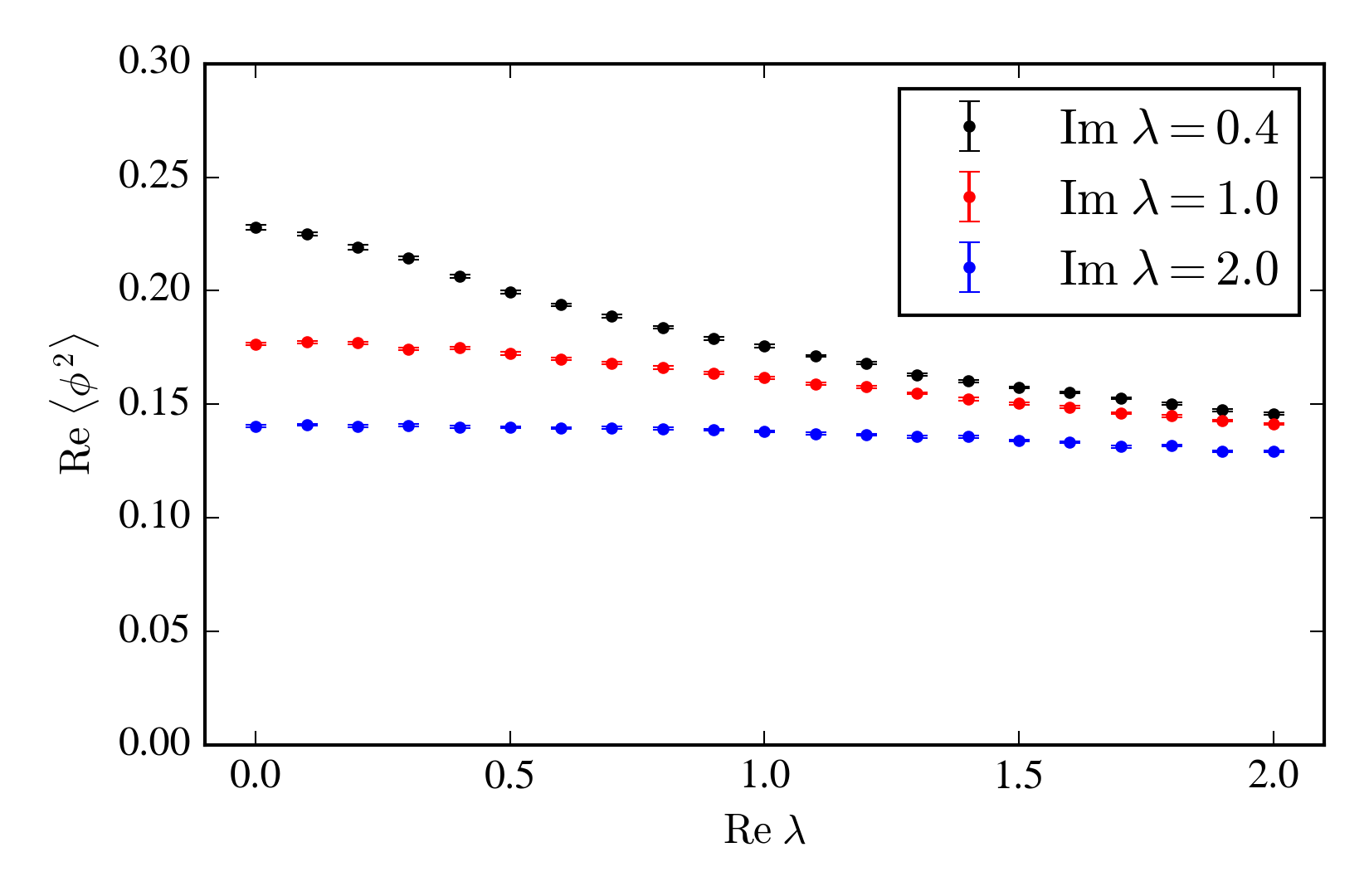}
\includegraphics[width=0.45\linewidth]{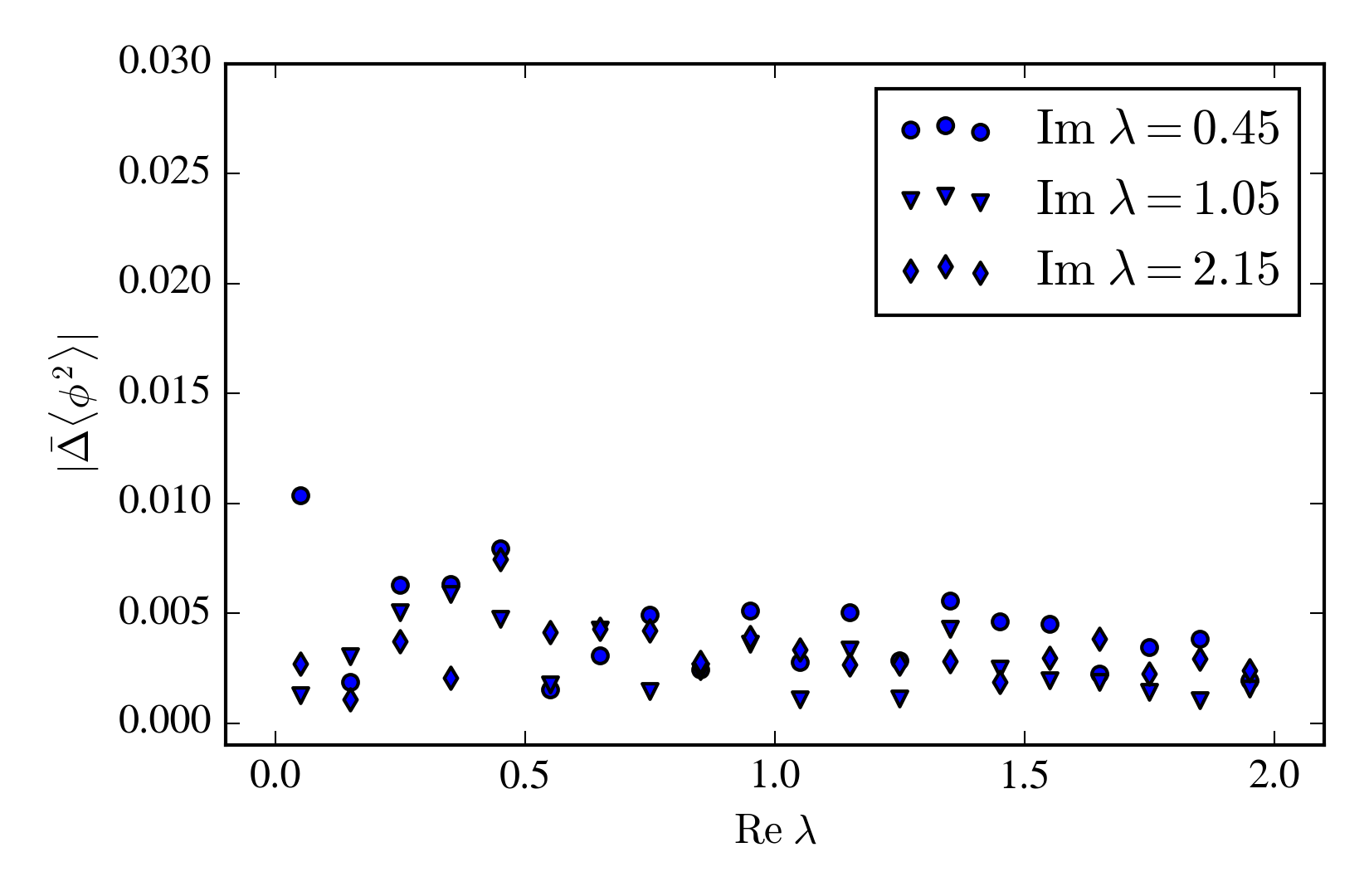}
\end{center}
\caption{The expectation value of $\langle \phi^2\rangle$, for $m^2=0.5$ on a $8\times8$ lattice. The left panel shows the Monte Carlo estimate of the expectation value generated from contours trained with $4\times10^5$ configurations as described in the text. The right panel shows the violation of the Cauchy-Riemann equations, approximated with finite differencing. Statistical errors, dwarfed by finite differencing errors, fit within the data points and are not shown. That the violation is small confirms that the Monte Carlo is indeed outputting the analytic continuation of the real-$\lambda$ results.\label{fig:2d-expect}}
\end{figure*}

The expectation value $\langle \phi^2\rangle$ is shown as a function of complex coupling $\lambda$ on the left side of Figure~\ref{fig:2d-expect}, for an $8 \times 8$ lattice with $m^2 = 0.5$.

As a check of the correctness of our implementation of the contour deformation method, we can verify that expectation values, as a function of $\lambda$, are in fact holomorphic. Provided the real-plane computation is trusted, the fact that the function $\langle \phi^2\rangle(\lambda)$ is holomorphic uniquely identifies that function. Of course this check can only be performed to within errors introduced by finite statistics and finite differencing. Figure~\ref{fig:2d-expect} shows the anti-holomorphic Wirtinger derivative $\frac{\partial}{\partial \bar z} \equiv \frac 1 2 \left(\frac{\partial}{\partial x} + i \frac{\partial}{\partial y}\right)$ of $\langle \phi^2 \rangle$, approximated by finite differencing according to
\begin{widetext}
\begin{equation}
\bar\Delta f(z)
\equiv
\frac 1 {4\delta} \left[
\frac{f(z_{++}) + f(z_{+-})}{2}
-
\frac{f(z_{-+}) + f(z_{--})}{2}
+
i
\left(
\frac{f(z_{++}) + f(z_{-+})}{2}
-
\frac{f(z_{+-}) + f(z_{--})}{2}
\right)
\right]
\text,
\end{equation}
\end{widetext}
where $z_{+\pm} = z + \delta \pm i\delta$ and $z_{-\pm} = z - \delta \pm i\delta$. We adopt a finite difference of $\delta = 0.1$. The small violation confirms that the Monte Carlo method on deformed contours is indeed outputting (within our ability to make measurements) a complex analytic function.

\section{Partition Function Zeros}\label{sec:zeros}
In~\cite{Lawrence:2021izu} it was argued that theories with polynomial actions typically have perfect contours, and several one-dimensional integrals were examined as examples. In the previous section, we gave numerical evidence of quantum mechanical path integrals that also have perfect contours. In this section, we explore a source of counter-examples: the zeros of a partition function.

To begin, consider the one-dimensional integral
\begin{equation}
I(\alpha) = \int_{\mathbb R} dz\;e^{-\alpha z^2 - z^4}
\text.
\end{equation}
Specific values of $\alpha$ were used as examples in~\cite{Lawrence:2021izu}, where perfect integration contours could be found. There are, however, values of $\alpha$ for which the best possible average phase becomes arbitrarily close to $0$. To see this, we must establish that:
\begin{enumerate}
\item For any $\epsilon > 0$, $|I(\alpha)| < \epsilon$ can be satisfied, and
\item For any fixed $\alpha$, the quenched partition function $I_Q(\alpha) = \int |e^{-\alpha z^2 - z^4}|$ has a non-zero lower bound.
\end{enumerate}
Given that both of these are true, we immediately obtain an bound on the average phase: $\langle \sigma\rangle = \frac{I}{I_Q} \le \frac{\epsilon}{B}$, where $B$ is the non-zero lower bound on the quenched partition function.

The first observation is an immediate consequence of the great Picard's theorem: since $I : \mathbb C \rightarrow \mathbb C$ has an essential singularity at infinity, all values in $\mathbb C$ with at most one exception are obtained. In fact, as is often visible numerically, regions where a partition function becomes arbitrarily small typically form (topological) balls around zeros of the partition function.

To establish a lower bound on the quenched partition function, consider the complex plane in polar coordinates $z = r e^{i \theta}$, and pick some fiducial radius $R$. Every contour $\gamma : \mathbb R \rightarrow \mathbb C$ either crosses within $R$, or stays always at $r > R$. In the former case, there must be a region $(t_-,t_+)$ such that $r([t_-,t_+]) = [R,R+1]$; that is, the contour must enter the annulus between radii $R$ and $R+1$ at coordinate $t_-$, and exit at $t_+$. Any such segment of the contour must be length at least $1$. Furthermore, there exists a positive $C_R$ such that $|e^{-\alpha z^2 - z^4}| > C_R$ for all $z$ in this annulus. As a result, the contribution of this section of the contour to the quenched partition function must be at least $C_R$.

In the latter case, the contour must cross through either $\theta \in (\frac\pi 8,\frac{3\pi}{8})$ or $\theta \in (-\frac\pi 8,-\frac{3\pi}{8})$ while remaining at radii greater than $R$---without loss of generality we may assume the former. In this ``unstable'' region, the integrand grows without bound as $r$ is made larger. As a result, there is again a positive $D_R$ such that $|e^{-\alpha z^2 - z^4}| > D_R$ everywhere in this region. The length of the portion of the contour in this region must be at least $\frac\pi 4 R$, and so we obtain a lower bound of $\frac\pi 4 R D_R$.

The lesser of the two bounds $C_R$ and $\frac\pi 4 R D_R$ provides our (far from tight) lower bound on the quenched partition function. In principle the choice of $R$ can be optimized (and made $\alpha$-dependent) to obtain a stronger bound, but this is not important here.
 
A non-zero lower bound on the quenched partition function is generically available in physical theories, although it may be difficult to compute a reasonably tight bound. For this reason, the proximity of partition function zeros may be seen as an obstacle to the existence of perfect contours, although not necessarily the only one.

With these considerations in mind, it is worth considering where zeros of the scalar field partition function occur. It is to be expected that there are such zeros; for example, when an external field source is added via a term $J \phi$ in the action, the existence of zeros in the complex plane of $J$ has long been known~\cite{PhysRevLett.30.931}.

Normalizing flows can be used to measure the partition function directly by a Monte Carlo, as demonstrated in the previous section; however, this does not immediately make them a practical tool for studying zeros of the partition function. A quick glance at Figure~\ref{fig:partition} determines that the presence or absence of zeros is not readily discernible. Primarily this is due to the fact that, as we have just seen, near zeros of the partition function the average phase becomes arbitrarily small. As a result, determining whether the partition function is zero or simply arbitrarily small at any given point is not possible.

Fortunately, the \emph{presence} of a zero (but not its absolute location) can be determined exclusively from Monte Carlo calculations done in a loop around the zero. 
Consider the following integral, taken over a loop $\gamma$ on the complex plane of $\lambda$:
\begin{equation}
I_\gamma = \frac 1 {2\pi i} \oint_\gamma d \lambda\; \frac{Z'(\lambda)}{Z(\lambda)}
\end{equation}
Provided that $\gamma$ is the boundary of some region on which $Z(\lambda)$ lacks any poles, $I_\gamma$ is an integer which counts (with multiplicity) the number of zeros of $Z(\lambda)$ in that region. The integral has an alternate interpretation as the change in the argument of $Z(\lambda)$ as the closed path $\gamma$ is followed. Thus, the number of zeros in a region appears as a winding number measured by walking around that region.

This allows us to study the zeros of the partition function without being able to perform Monte Carlo calculations at or near them. In the two dimensional case studied in Section~\ref{sec:2d}, we have access to only a contour deformation, making the determination of $Z$ itself more difficult. However, the argument of $Z$ (i.e. the imaginary part of the free energy) appears as argument of the average phase; that is, only the magnitude of $Z$ is difficult to determine via Monte Carlo. Since $Z(\lambda)$ is holomorphic for $\lambda \ne 0$, it is possible to locate zeros purely by inspecting the phase.

\begin{figure}
\centering\includegraphics[width=0.94\linewidth]{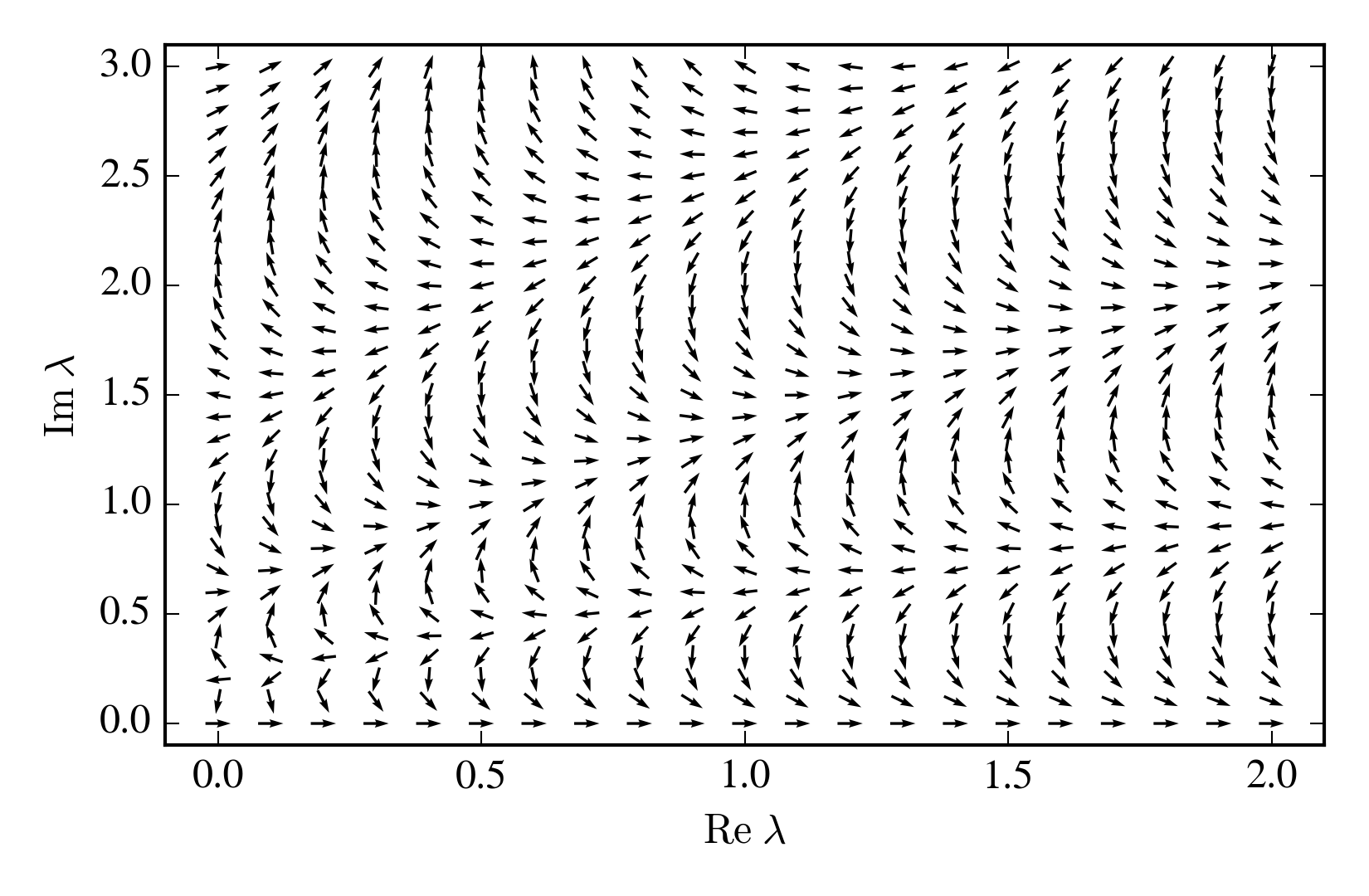}
\caption{Phases of the partition function for $m^2=0.5$ on an $8\times 8$ lattice, on the complex-$\lambda$ plane. Since the partition function provably has no singularities away from $\lambda = 0$ (a branch point), the number of zeros in any region may be determined by counting the winding number of a curve around that region.\label{fig:phases}}
\end{figure}

Figure~\ref{fig:phases} plots the phase of the partition function as a function of complex coupling $\lambda$, for the first quadrant of the complex plane. The presence or absence of zeros in any given region can be read off of the plot by counting the winding number of a loop around that region. It is thus clearly visible from this plot that there are no zeros in the first quadrant of the complex plane (at least for $|\lambda|\lesssim 2$). Since the magnitude of the partition function is symmetric under complex conjugation of the coupling---$Z(\bar \lambda) = \overline{Z(\lambda)}$-- it is also true that there are no zeros in the corresponding part of the fourth quadrant.

\section{Discussion}\label{sec:discussion}
Using complex $\phi^4$ coupling as a test case, we have demonstrated the utility of contour deformation-based approaches, including complex normalizing flows, in the study of non-Hermitian lattice systems of scalar fields. It appears likely that, as argued in~\cite{Lawrence:2021izu}, perfect integration contours exist for most values of the coupling.

Section~\ref{sec:zeros} outlined a connection between zeros of the partition function and contour deformations for treating the sign problem. Proximity to a zero of the partition function creates an obstacle to the existence of perfect contour deformations, and around any such zero there is guaranteed to be a region where no perfect contour exists. When the zeros occur on the complex plane of an external field, they are known as Lee-Yang zeros~\cite{Lee:1952ig}, and their existence and behavior has been long studied in scalar field theory~\cite{PhysRevLett.30.931} and recently in~\cite{Pawlowski:2022rdn} via a flow-based density-of-states approach. In the complex plane of classical temperature ($\hbar$ in a quantum path integral), the zeros are referred to as Fisher zeros~\cite{fisher1965nature}. The significance of partition function zeros for contour deformation-based studies of lattice systems in physical regimes---where the Hamiltonian is Hermitian and $|Z| > 0$---remains to be determined.

We have not in this paper discussed phase transitions---either the second-order transitions that characterize the continuum limit of the scalar field theory being studied here, or first-order transitions that occur in many other models. In the context of the Lefschetz-thimble approach (the historical inspiration for both contour deformations and complex normalizing flows), the role of phase transitions and their relationship to the thimbles has been well studied~\cite{Kanazawa:2014qma,Tanizaki:2015gpl}. A careful study of the relationship between phase transitions and optimal contours is desirable.

An intricate set of tradeoffs exists between the two approaches used in this paper. Directly optimizing contour deformations requires a laborious Monte Carlo to be performed at every step of the gradient descent. When optimizing a complex normalizing flow, the sampling is made cheap---after all, the original (real) formulation of normalizing flows is typically used as a technique for dramatically accelerating Monte Carlo sampling. However, learning an accurate normalizing flow is considerably more effort; as observed in Section~\ref{sec:2d}, linear contour deformations are already quite effective in treating the complex-coupling sign problem, whereas an approximate normalizing flow must be strongly nonlinear even when the coupling is real. Moreover, as mentioned in~\cite{Lawrence:2021izu}, a normalizing flow cannot be effectively trained in regimes where an unresolved severe sign problem exists. Finally, the procedures for training a contour deformation (and the conditions under which the trained contour respects the conditions of Cauchy's integral theorem) are well understood; training normalizing flows correctly currently requires some finesse.

Most immediately, the work done here suggests the practicality of performing lattice studies of the Lee-Yang or Fisher zeros of systems with scalar degrees of freedom. The success of contour deformation methods in treating \emph{homogeneous} complex couplings also bodes well for lattice studies of real-time dynamics, which (making use of the lattice Schwinger-Keldysh formalism) may be thought of as possessing particular inhomogeneous complex couplings.

\begin{acknowledgments}
The authors are grateful to Brian Lawrence for pointing out the existence of integration contours that cannot be parametrized by the real plane, and suggesting the method of studying the zeros of the partition function. Y.Y.~appreciates the hospitality of Paul Romatschke and the University of Colorado at Boulder.

S.L.~is supported by the U.S.~Department of Energy under Contract
No.~DE-SC0017905. H.O.~and Y.Y.~are supported by the U.S.~Department of Energy under
Contract No.~DE-FG02-93ER-40762.
\end{acknowledgments}
\bibliographystyle{apsrev4-2}
\bibliography{Self,References}

\begin{thebibliography}{27}%
\makeatletter
\providecommand \@ifxundefined [1]{%
 \@ifx{#1\undefined}
}%
\providecommand \@ifnum [1]{%
 \ifnum #1\expandafter \@firstoftwo
 \else \expandafter \@secondoftwo
 \fi
}%
\providecommand \@ifx [1]{%
 \ifx #1\expandafter \@firstoftwo
 \else \expandafter \@secondoftwo
 \fi
}%
\providecommand \natexlab [1]{#1}%
\providecommand \enquote  [1]{``#1''}%
\providecommand \bibnamefont  [1]{#1}%
\providecommand \bibfnamefont [1]{#1}%
\providecommand \citenamefont [1]{#1}%
\providecommand \href@noop [0]{\@secondoftwo}%
\providecommand \href [0]{\begingroup \@sanitize@url \@href}%
\providecommand \@href[1]{\@@startlink{#1}\@@href}%
\providecommand \@@href[1]{\endgroup#1\@@endlink}%
\providecommand \@sanitize@url [0]{\catcode `\\12\catcode `\$12\catcode
  `\&12\catcode `\#12\catcode `\^12\catcode `\_12\catcode `\%12\relax}%
\providecommand \@@startlink[1]{}%
\providecommand \@@endlink[0]{}%
\providecommand \url  [0]{\begingroup\@sanitize@url \@url }%
\providecommand \@url [1]{\endgroup\@href {#1}{\urlprefix }}%
\providecommand \urlprefix  [0]{URL }%
\providecommand \Eprint [0]{\href }%
\providecommand \doibase [0]{https://doi.org/}%
\providecommand \selectlanguage [0]{\@gobble}%
\providecommand \bibinfo  [0]{\@secondoftwo}%
\providecommand \bibfield  [0]{\@secondoftwo}%
\providecommand \translation [1]{[#1]}%
\providecommand \BibitemOpen [0]{}%
\providecommand \bibitemStop [0]{}%
\providecommand \bibitemNoStop [0]{.\EOS\space}%
\providecommand \EOS [0]{\spacefactor3000\relax}%
\providecommand \BibitemShut  [1]{\csname bibitem#1\endcsname}%
\let\auto@bib@innerbib\@empty
\bibitem [{\citenamefont {Alexandru}\ \emph
  {et~al.}(2016{\natexlab{a}})\citenamefont {Alexandru}, \citenamefont {Basar},
  \citenamefont {Bedaque}, \citenamefont {Vartak},\ and\ \citenamefont
  {Warrington}}]{Alexandru:2016gsd}%
  \BibitemOpen
  \bibfield  {author} {\bibinfo {author} {\bibfnamefont {A.}~\bibnamefont
  {Alexandru}}, \bibinfo {author} {\bibfnamefont {G.}~\bibnamefont {Basar}},
  \bibinfo {author} {\bibfnamefont {P.~F.}\ \bibnamefont {Bedaque}}, \bibinfo
  {author} {\bibfnamefont {S.}~\bibnamefont {Vartak}},\ and\ \bibinfo {author}
  {\bibfnamefont {N.~C.}\ \bibnamefont {Warrington}},\ }\href
  {https://doi.org/10.1103/PhysRevLett.117.081602} {\bibfield  {journal}
  {\bibinfo  {journal} {Phys. Rev. Lett.}\ }\textbf {\bibinfo {volume} {117}},\
  \bibinfo {pages} {081602} (\bibinfo {year} {2016}{\natexlab{a}})},\ \Eprint
  {https://arxiv.org/abs/1605.08040} {arXiv:1605.08040 [hep-lat]} \BibitemShut
  {NoStop}%
\bibitem [{\citenamefont {Alexandru}\ \emph
  {et~al.}(2017{\natexlab{a}})\citenamefont {Alexandru}, \citenamefont {Basar},
  \citenamefont {Bedaque},\ and\ \citenamefont {Ridgway}}]{Alexandru:2017lqr}%
  \BibitemOpen
  \bibfield  {author} {\bibinfo {author} {\bibfnamefont {A.}~\bibnamefont
  {Alexandru}}, \bibinfo {author} {\bibfnamefont {G.}~\bibnamefont {Basar}},
  \bibinfo {author} {\bibfnamefont {P.~F.}\ \bibnamefont {Bedaque}},\ and\
  \bibinfo {author} {\bibfnamefont {G.~W.}\ \bibnamefont {Ridgway}},\ }\href
  {https://doi.org/10.1103/PhysRevD.95.114501} {\bibfield  {journal} {\bibinfo
  {journal} {Phys. Rev. D}\ }\textbf {\bibinfo {volume} {95}},\ \bibinfo
  {pages} {114501} (\bibinfo {year} {2017}{\natexlab{a}})},\ \Eprint
  {https://arxiv.org/abs/1704.06404} {arXiv:1704.06404 [hep-lat]} \BibitemShut
  {NoStop}%
\bibitem [{\citenamefont {Witten}(2011)}]{Witten:2010cx}%
  \BibitemOpen
  \bibfield  {author} {\bibinfo {author} {\bibfnamefont {E.}~\bibnamefont
  {Witten}},\ }\href@noop {} {\bibfield  {journal} {\bibinfo  {journal} {AMS/IP
  Stud. Adv. Math.}\ }\textbf {\bibinfo {volume} {50}},\ \bibinfo {pages} {347}
  (\bibinfo {year} {2011})},\ \Eprint {https://arxiv.org/abs/1001.2933}
  {arXiv:1001.2933 [hep-th]} \BibitemShut {NoStop}%
\bibitem [{\citenamefont {Cristoforetti}\ \emph {et~al.}(2012)\citenamefont
  {Cristoforetti}, \citenamefont {Di~Renzo},\ and\ \citenamefont
  {Scorzato}}]{Cristoforetti:2012su}%
  \BibitemOpen
  \bibfield  {author} {\bibinfo {author} {\bibfnamefont {M.}~\bibnamefont
  {Cristoforetti}}, \bibinfo {author} {\bibfnamefont {F.}~\bibnamefont
  {Di~Renzo}},\ and\ \bibinfo {author} {\bibfnamefont {L.}~\bibnamefont
  {Scorzato}} (\bibinfo {collaboration} {AuroraScience}),\ }\href
  {https://doi.org/10.1103/PhysRevD.86.074506} {\bibfield  {journal} {\bibinfo
  {journal} {Phys. Rev. D}\ }\textbf {\bibinfo {volume} {86}},\ \bibinfo
  {pages} {074506} (\bibinfo {year} {2012})},\ \Eprint
  {https://arxiv.org/abs/1205.3996} {arXiv:1205.3996 [hep-lat]} \BibitemShut
  {NoStop}%
\bibitem [{\citenamefont {Alexandru}\ \emph
  {et~al.}(2016{\natexlab{b}})\citenamefont {Alexandru}, \citenamefont
  {Basar},\ and\ \citenamefont {Bedaque}}]{Alexandru:2015xva}%
  \BibitemOpen
  \bibfield  {author} {\bibinfo {author} {\bibfnamefont {A.}~\bibnamefont
  {Alexandru}}, \bibinfo {author} {\bibfnamefont {G.}~\bibnamefont {Basar}},\
  and\ \bibinfo {author} {\bibfnamefont {P.}~\bibnamefont {Bedaque}},\ }\href
  {https://doi.org/10.1103/PhysRevD.93.014504} {\bibfield  {journal} {\bibinfo
  {journal} {Phys. Rev. D}\ }\textbf {\bibinfo {volume} {93}},\ \bibinfo
  {pages} {014504} (\bibinfo {year} {2016}{\natexlab{b}})},\ \Eprint
  {https://arxiv.org/abs/1510.03258} {arXiv:1510.03258 [hep-lat]} \BibitemShut
  {NoStop}%
\bibitem [{\citenamefont {Alexandru}\ \emph
  {et~al.}(2016{\natexlab{c}})\citenamefont {Alexandru}, \citenamefont {Basar},
  \citenamefont {Bedaque}, \citenamefont {Ridgway},\ and\ \citenamefont
  {Warrington}}]{Alexandru:2015sua}%
  \BibitemOpen
  \bibfield  {author} {\bibinfo {author} {\bibfnamefont {A.}~\bibnamefont
  {Alexandru}}, \bibinfo {author} {\bibfnamefont {G.}~\bibnamefont {Basar}},
  \bibinfo {author} {\bibfnamefont {P.~F.}\ \bibnamefont {Bedaque}}, \bibinfo
  {author} {\bibfnamefont {G.~W.}\ \bibnamefont {Ridgway}},\ and\ \bibinfo
  {author} {\bibfnamefont {N.~C.}\ \bibnamefont {Warrington}},\ }\href
  {https://doi.org/10.1007/JHEP05(2016)053} {\bibfield  {journal} {\bibinfo
  {journal} {JHEP}\ }\textbf {\bibinfo {volume} {05}},\ \bibinfo {pages}
  {053}},\ \Eprint {https://arxiv.org/abs/1512.08764} {arXiv:1512.08764
  [hep-lat]} \BibitemShut {NoStop}%
\bibitem [{\citenamefont {Lawrence}(2018)}]{Lawrence:2018mve}%
  \BibitemOpen
  \bibfield  {author} {\bibinfo {author} {\bibfnamefont {S.}~\bibnamefont
  {Lawrence}},\ }\href {https://doi.org/10.22323/1.334.0149} {\bibfield
  {journal} {\bibinfo  {journal} {PoS}\ }\textbf {\bibinfo {volume}
  {LATTICE2018}},\ \bibinfo {pages} {149} (\bibinfo {year} {2018})},\ \Eprint
  {https://arxiv.org/abs/1810.06529} {arXiv:1810.06529 [hep-lat]} \BibitemShut
  {NoStop}%
\bibitem [{\citenamefont {Alexandru}\ \emph {et~al.}(2020)\citenamefont
  {Alexandru}, \citenamefont {Basar}, \citenamefont {Bedaque},\ and\
  \citenamefont {Warrington}}]{Alexandru:2020wrj}%
  \BibitemOpen
  \bibfield  {author} {\bibinfo {author} {\bibfnamefont {A.}~\bibnamefont
  {Alexandru}}, \bibinfo {author} {\bibfnamefont {G.}~\bibnamefont {Basar}},
  \bibinfo {author} {\bibfnamefont {P.~F.}\ \bibnamefont {Bedaque}},\ and\
  \bibinfo {author} {\bibfnamefont {N.~C.}\ \bibnamefont {Warrington}},\
  }\href@noop {} {\bibinfo {title} {{Complex Paths Around The Sign Problem}}}
  (\bibinfo {year} {2020}),\ \Eprint {https://arxiv.org/abs/2007.05436}
  {arXiv:2007.05436 [hep-lat]} \BibitemShut {NoStop}%
\bibitem [{\citenamefont {Alexandru}\ \emph
  {et~al.}(2017{\natexlab{b}})\citenamefont {Alexandru}, \citenamefont
  {Bedaque}, \citenamefont {Lamm},\ and\ \citenamefont
  {Lawrence}}]{Alexandru:2017czx}%
  \BibitemOpen
  \bibfield  {author} {\bibinfo {author} {\bibfnamefont {A.}~\bibnamefont
  {Alexandru}}, \bibinfo {author} {\bibfnamefont {P.~F.}\ \bibnamefont
  {Bedaque}}, \bibinfo {author} {\bibfnamefont {H.}~\bibnamefont {Lamm}},\ and\
  \bibinfo {author} {\bibfnamefont {S.}~\bibnamefont {Lawrence}},\ }\href
  {https://doi.org/10.1103/PhysRevD.96.094505} {\bibfield  {journal} {\bibinfo
  {journal} {Phys. Rev. D}\ }\textbf {\bibinfo {volume} {96}},\ \bibinfo
  {pages} {094505} (\bibinfo {year} {2017}{\natexlab{b}})},\ \Eprint
  {https://arxiv.org/abs/1709.01971} {arXiv:1709.01971 [hep-lat]} \BibitemShut
  {NoStop}%
\bibitem [{\citenamefont {Mori}\ \emph {et~al.}(2017)\citenamefont {Mori},
  \citenamefont {Kashiwa},\ and\ \citenamefont {Ohnishi}}]{Mori:2017pne}%
  \BibitemOpen
  \bibfield  {author} {\bibinfo {author} {\bibfnamefont {Y.}~\bibnamefont
  {Mori}}, \bibinfo {author} {\bibfnamefont {K.}~\bibnamefont {Kashiwa}},\ and\
  \bibinfo {author} {\bibfnamefont {A.}~\bibnamefont {Ohnishi}},\ }\href
  {https://doi.org/10.1103/PhysRevD.96.111501} {\bibfield  {journal} {\bibinfo
  {journal} {Phys. Rev. D}\ }\textbf {\bibinfo {volume} {96}},\ \bibinfo
  {pages} {111501} (\bibinfo {year} {2017})},\ \Eprint
  {https://arxiv.org/abs/1705.05605} {arXiv:1705.05605 [hep-lat]} \BibitemShut
  {NoStop}%
\bibitem [{\citenamefont {Alexandru}\ \emph
  {et~al.}(2018{\natexlab{a}})\citenamefont {Alexandru}, \citenamefont
  {Bedaque}, \citenamefont {Lamm},\ and\ \citenamefont
  {Lawrence}}]{Alexandru:2018fqp}%
  \BibitemOpen
  \bibfield  {author} {\bibinfo {author} {\bibfnamefont {A.}~\bibnamefont
  {Alexandru}}, \bibinfo {author} {\bibfnamefont {P.~F.}\ \bibnamefont
  {Bedaque}}, \bibinfo {author} {\bibfnamefont {H.}~\bibnamefont {Lamm}},\ and\
  \bibinfo {author} {\bibfnamefont {S.}~\bibnamefont {Lawrence}},\ }\href
  {https://doi.org/10.1103/PhysRevD.97.094510} {\bibfield  {journal} {\bibinfo
  {journal} {Phys. Rev. D}\ }\textbf {\bibinfo {volume} {97}},\ \bibinfo
  {pages} {094510} (\bibinfo {year} {2018}{\natexlab{a}})},\ \Eprint
  {https://arxiv.org/abs/1804.00697} {arXiv:1804.00697 [hep-lat]} \BibitemShut
  {NoStop}%
\bibitem [{\citenamefont {Giordano}\ \emph {et~al.}(2022)\citenamefont
  {Giordano}, \citenamefont {Kapas}, \citenamefont {Katz}, \citenamefont
  {Pasztor},\ and\ \citenamefont {Tulipant}}]{Giordano:2022miv}%
  \BibitemOpen
  \bibfield  {author} {\bibinfo {author} {\bibfnamefont {M.}~\bibnamefont
  {Giordano}}, \bibinfo {author} {\bibfnamefont {K.}~\bibnamefont {Kapas}},
  \bibinfo {author} {\bibfnamefont {S.~D.}\ \bibnamefont {Katz}}, \bibinfo
  {author} {\bibfnamefont {A.}~\bibnamefont {Pasztor}},\ and\ \bibinfo {author}
  {\bibfnamefont {Z.}~\bibnamefont {Tulipant}},\ }\href@noop {} {\bibinfo
  {title} {{Exponential reduction of the sign problem at finite density in the
  2+1D XY model via contour deformations}}} (\bibinfo {year} {2022}),\ \Eprint
  {https://arxiv.org/abs/2202.07561} {arXiv:2202.07561 [hep-lat]} \BibitemShut
  {NoStop}%
\bibitem [{\citenamefont {Lawrence}\ and\ \citenamefont
  {Yamauchi}(2021)}]{Lawrence:2021izu}%
  \BibitemOpen
  \bibfield  {author} {\bibinfo {author} {\bibfnamefont {S.}~\bibnamefont
  {Lawrence}}\ and\ \bibinfo {author} {\bibfnamefont {Y.}~\bibnamefont
  {Yamauchi}},\ }\href {https://doi.org/10.1103/PhysRevD.103.114509} {\bibfield
   {journal} {\bibinfo  {journal} {Phys. Rev. D}\ }\textbf {\bibinfo {volume}
  {103}},\ \bibinfo {pages} {114509} (\bibinfo {year} {2021})},\ \Eprint
  {https://arxiv.org/abs/2101.05755} {arXiv:2101.05755 [hep-lat]} \BibitemShut
  {NoStop}%
\bibitem [{\citenamefont {Hackett}\ \emph {et~al.}(2021)\citenamefont
  {Hackett}, \citenamefont {Hsieh}, \citenamefont {Albergo}, \citenamefont
  {Boyda}, \citenamefont {Chen}, \citenamefont {Chen}, \citenamefont {Cranmer},
  \citenamefont {Kanwar},\ and\ \citenamefont {Shanahan}}]{Hackett:2021idh}%
  \BibitemOpen
  \bibfield  {author} {\bibinfo {author} {\bibfnamefont {D.~C.}\ \bibnamefont
  {Hackett}}, \bibinfo {author} {\bibfnamefont {C.-C.}\ \bibnamefont {Hsieh}},
  \bibinfo {author} {\bibfnamefont {M.~S.}\ \bibnamefont {Albergo}}, \bibinfo
  {author} {\bibfnamefont {D.}~\bibnamefont {Boyda}}, \bibinfo {author}
  {\bibfnamefont {J.-W.}\ \bibnamefont {Chen}}, \bibinfo {author}
  {\bibfnamefont {K.-F.}\ \bibnamefont {Chen}}, \bibinfo {author}
  {\bibfnamefont {K.}~\bibnamefont {Cranmer}}, \bibinfo {author} {\bibfnamefont
  {G.}~\bibnamefont {Kanwar}},\ and\ \bibinfo {author} {\bibfnamefont {P.~E.}\
  \bibnamefont {Shanahan}},\ }\href@noop {} {\bibinfo {title} {{Flow-based
  sampling for multimodal distributions in lattice field theory}}} (\bibinfo
  {year} {2021}),\ \Eprint {https://arxiv.org/abs/2107.00734} {arXiv:2107.00734
  [hep-lat]} \BibitemShut {NoStop}%
\bibitem [{\citenamefont {Kanwar}\ \emph {et~al.}(2020)\citenamefont {Kanwar},
  \citenamefont {Albergo}, \citenamefont {Boyda}, \citenamefont {Cranmer},
  \citenamefont {Hackett}, \citenamefont {Racani\`ere}, \citenamefont
  {Rezende},\ and\ \citenamefont {Shanahan}}]{Kanwar:2020xzo}%
  \BibitemOpen
  \bibfield  {author} {\bibinfo {author} {\bibfnamefont {G.}~\bibnamefont
  {Kanwar}}, \bibinfo {author} {\bibfnamefont {M.~S.}\ \bibnamefont {Albergo}},
  \bibinfo {author} {\bibfnamefont {D.}~\bibnamefont {Boyda}}, \bibinfo
  {author} {\bibfnamefont {K.}~\bibnamefont {Cranmer}}, \bibinfo {author}
  {\bibfnamefont {D.~C.}\ \bibnamefont {Hackett}}, \bibinfo {author}
  {\bibfnamefont {S.}~\bibnamefont {Racani\`ere}}, \bibinfo {author}
  {\bibfnamefont {D.~J.}\ \bibnamefont {Rezende}},\ and\ \bibinfo {author}
  {\bibfnamefont {P.~E.}\ \bibnamefont {Shanahan}},\ }\href
  {https://doi.org/10.1103/PhysRevLett.125.121601} {\bibfield  {journal}
  {\bibinfo  {journal} {Phys. Rev. Lett.}\ }\textbf {\bibinfo {volume} {125}},\
  \bibinfo {pages} {121601} (\bibinfo {year} {2020})},\ \Eprint
  {https://arxiv.org/abs/2003.06413} {arXiv:2003.06413 [hep-lat]} \BibitemShut
  {NoStop}%
\bibitem [{\citenamefont {Albergo}\ \emph {et~al.}(2019)\citenamefont
  {Albergo}, \citenamefont {Kanwar},\ and\ \citenamefont
  {Shanahan}}]{Albergo:2019eim}%
  \BibitemOpen
  \bibfield  {author} {\bibinfo {author} {\bibfnamefont {M.~S.}\ \bibnamefont
  {Albergo}}, \bibinfo {author} {\bibfnamefont {G.}~\bibnamefont {Kanwar}},\
  and\ \bibinfo {author} {\bibfnamefont {P.~E.}\ \bibnamefont {Shanahan}},\
  }\href {https://doi.org/10.1103/PhysRevD.100.034515} {\bibfield  {journal}
  {\bibinfo  {journal} {Phys. Rev. D}\ }\textbf {\bibinfo {volume} {100}},\
  \bibinfo {pages} {034515} (\bibinfo {year} {2019})},\ \Eprint
  {https://arxiv.org/abs/1904.12072} {arXiv:1904.12072 [hep-lat]} \BibitemShut
  {NoStop}%
\bibitem [{\citenamefont {Albergo}\ \emph {et~al.}(2021)\citenamefont
  {Albergo}, \citenamefont {Kanwar}, \citenamefont {Racani\`ere}, \citenamefont
  {Rezende}, \citenamefont {Urban}, \citenamefont {Boyda}, \citenamefont
  {Cranmer}, \citenamefont {Hackett},\ and\ \citenamefont
  {Shanahan}}]{Albergo:2021bna}%
  \BibitemOpen
  \bibfield  {author} {\bibinfo {author} {\bibfnamefont {M.~S.}\ \bibnamefont
  {Albergo}}, \bibinfo {author} {\bibfnamefont {G.}~\bibnamefont {Kanwar}},
  \bibinfo {author} {\bibfnamefont {S.}~\bibnamefont {Racani\`ere}}, \bibinfo
  {author} {\bibfnamefont {D.~J.}\ \bibnamefont {Rezende}}, \bibinfo {author}
  {\bibfnamefont {J.~M.}\ \bibnamefont {Urban}}, \bibinfo {author}
  {\bibfnamefont {D.}~\bibnamefont {Boyda}}, \bibinfo {author} {\bibfnamefont
  {K.}~\bibnamefont {Cranmer}}, \bibinfo {author} {\bibfnamefont {D.~C.}\
  \bibnamefont {Hackett}},\ and\ \bibinfo {author} {\bibfnamefont {P.~E.}\
  \bibnamefont {Shanahan}},\ }\href
  {https://doi.org/10.1103/PhysRevD.104.114507} {\bibfield  {journal} {\bibinfo
   {journal} {Phys. Rev. D}\ }\textbf {\bibinfo {volume} {104}},\ \bibinfo
  {pages} {114507} (\bibinfo {year} {2021})},\ \Eprint
  {https://arxiv.org/abs/2106.05934} {arXiv:2106.05934 [hep-lat]} \BibitemShut
  {NoStop}%
\bibitem [{cod()}]{code}%
  \BibitemOpen
  \href {https://gitlab.com/s.lawrence/cauchy} {}\bibinfo {howpublished}
  {https://gitlab.com/s.lawrence/cauchy}\BibitemShut {NoStop}%
\bibitem [{\citenamefont {Lawrence}(2020)}]{Lawrence:2020irw}%
  \BibitemOpen
  \bibfield  {author} {\bibinfo {author} {\bibfnamefont {S.}~\bibnamefont
  {Lawrence}},\ }\emph {\bibinfo {title} {{Sign Problems in Quantum Field
  Theory: Classical and Quantum Approaches}}},\ \href@noop {} {Ph.D. thesis},\
  \bibinfo  {school} {Maryland U.} (\bibinfo {year} {2020}),\ \Eprint
  {https://arxiv.org/abs/2006.03683} {arXiv:2006.03683 [hep-lat]} \BibitemShut
  {NoStop}%
\bibitem [{\citenamefont {Kingma}\ and\ \citenamefont
  {Ba}(2014)}]{kingma2014adam}%
  \BibitemOpen
  \bibfield  {author} {\bibinfo {author} {\bibfnamefont {D.~P.}\ \bibnamefont
  {Kingma}}\ and\ \bibinfo {author} {\bibfnamefont {J.}~\bibnamefont {Ba}},\
  }\href@noop {} {\bibfield  {journal} {\bibinfo  {journal} {arXiv preprint
  arXiv:1412.6980}\ } (\bibinfo {year} {2014})}\BibitemShut {NoStop}%
\bibitem [{\citenamefont {Alexandru}\ \emph
  {et~al.}(2018{\natexlab{b}})\citenamefont {Alexandru}, \citenamefont
  {Bedaque}, \citenamefont {Lamm}, \citenamefont {Lawrence},\ and\
  \citenamefont {Warrington}}]{Alexandru:2018ddf}%
  \BibitemOpen
  \bibfield  {author} {\bibinfo {author} {\bibfnamefont {A.}~\bibnamefont
  {Alexandru}}, \bibinfo {author} {\bibfnamefont {P.~F.}\ \bibnamefont
  {Bedaque}}, \bibinfo {author} {\bibfnamefont {H.}~\bibnamefont {Lamm}},
  \bibinfo {author} {\bibfnamefont {S.}~\bibnamefont {Lawrence}},\ and\
  \bibinfo {author} {\bibfnamefont {N.~C.}\ \bibnamefont {Warrington}},\ }\href
  {https://doi.org/10.1103/PhysRevLett.121.191602} {\bibfield  {journal}
  {\bibinfo  {journal} {Phys. Rev. Lett.}\ }\textbf {\bibinfo {volume} {121}},\
  \bibinfo {pages} {191602} (\bibinfo {year} {2018}{\natexlab{b}})},\ \Eprint
  {https://arxiv.org/abs/1808.09799} {arXiv:1808.09799 [hep-lat]} \BibitemShut
  {NoStop}%
\bibitem [{\citenamefont {Simon}\ and\ \citenamefont
  {Griffiths}(1973)}]{PhysRevLett.30.931}%
  \BibitemOpen
  \bibfield  {author} {\bibinfo {author} {\bibfnamefont {B.}~\bibnamefont
  {Simon}}\ and\ \bibinfo {author} {\bibfnamefont {R.~B.}\ \bibnamefont
  {Griffiths}},\ }\href {https://doi.org/10.1103/PhysRevLett.30.931} {\bibfield
   {journal} {\bibinfo  {journal} {Phys. Rev. Lett.}\ }\textbf {\bibinfo
  {volume} {30}},\ \bibinfo {pages} {931} (\bibinfo {year} {1973})}\BibitemShut
  {NoStop}%
\bibitem [{\citenamefont {Lee}\ and\ \citenamefont {Yang}(1952)}]{Lee:1952ig}%
  \BibitemOpen
  \bibfield  {author} {\bibinfo {author} {\bibfnamefont {T.~D.}\ \bibnamefont
  {Lee}}\ and\ \bibinfo {author} {\bibfnamefont {C.-N.}\ \bibnamefont {Yang}},\
  }\href {https://doi.org/10.1103/PhysRev.87.410} {\bibfield  {journal}
  {\bibinfo  {journal} {Phys. Rev.}\ }\textbf {\bibinfo {volume} {87}},\
  \bibinfo {pages} {410} (\bibinfo {year} {1952})}\BibitemShut {NoStop}%
\bibitem [{\citenamefont {Pawlowski}\ and\ \citenamefont
  {Urban}(2022)}]{Pawlowski:2022rdn}%
  \BibitemOpen
  \bibfield  {author} {\bibinfo {author} {\bibfnamefont {J.~M.}\ \bibnamefont
  {Pawlowski}}\ and\ \bibinfo {author} {\bibfnamefont {J.~M.}\ \bibnamefont
  {Urban}},\ }\href@noop {} {\bibinfo {title} {{Flow-based density of states
  for complex actions}}} (\bibinfo {year} {2022}),\ \Eprint
  {https://arxiv.org/abs/2203.01243} {arXiv:2203.01243 [hep-lat]} \BibitemShut
  {NoStop}%
\bibitem [{\citenamefont {Fisher}(1965)}]{fisher1965nature}%
  \BibitemOpen
  \bibfield  {author} {\bibinfo {author} {\bibfnamefont {M.~E.}\ \bibnamefont
  {Fisher}},\ }\href@noop {} {\emph {\bibinfo {title} {The nature of critical
  points}}}\ (\bibinfo  {publisher} {University of Colorado Press},\ \bibinfo
  {year} {1965})\BibitemShut {NoStop}%
\bibitem [{\citenamefont {Kanazawa}\ and\ \citenamefont
  {Tanizaki}(2015)}]{Kanazawa:2014qma}%
  \BibitemOpen
  \bibfield  {author} {\bibinfo {author} {\bibfnamefont {T.}~\bibnamefont
  {Kanazawa}}\ and\ \bibinfo {author} {\bibfnamefont {Y.}~\bibnamefont
  {Tanizaki}},\ }\href {https://doi.org/10.1007/JHEP03(2015)044} {\bibfield
  {journal} {\bibinfo  {journal} {JHEP}\ }\textbf {\bibinfo {volume} {03}},\
  \bibinfo {pages} {044}},\ \Eprint {https://arxiv.org/abs/1412.2802}
  {arXiv:1412.2802 [hep-th]} \BibitemShut {NoStop}%
\bibitem [{\citenamefont {Tanizaki}(2015)}]{Tanizaki:2015gpl}%
  \BibitemOpen
  \bibfield  {author} {\bibinfo {author} {\bibfnamefont {Y.}~\bibnamefont
  {Tanizaki}},\ }\emph {\bibinfo {title} {{Study on sign problem via
  Lefschetz-thimble path integral}}},\ \href
  {https://doi.org/10.15083/00073296} {Ph.D. thesis},\ \bibinfo  {school}
  {Tokyo U.} (\bibinfo {year} {2015})\BibitemShut {NoStop}%
\end{thebibliography}%
\end{document}